\documentclass[11pt,twoside]{article}

\usepackage{asp2006}
\usepackage{epsfig}
\usepackage{amsmath,amscd,amsthm,amsfonts}
\usepackage{amsgen,amstext,amsbsy,amsopn}

\begin{document}
\title{Vortices and Angular Momentum in Bose-Einstein-Condensed Cold Dark Matter Halos}   
\author{Tanja Rindler-Daller$^1$ and Paul R. Shapiro$^2$}   
\affil{$^1$ Institut f\"ur Theoretische Physik, \\Universit\"at zu K\"oln, 50937 K\"oln, Germany\\
$^2$ Department of Astronomy \& Texas Cosmology Center, \\The University of Texas at Austin, 78712 Austin, USA}    

\begin{abstract}
If cold dark matter elementary particles form a Bose-Einstein
condensate, their superfluidity may distinguish them from other
forms of cold dark matter, including creation of quantum vortices.
We demonstrate here that such vortices are favoured in
strongly-coupled condensates, while this is not the case for axions,
which are generally presumed to form a Bose-Einstein condensate but
are effectively non-interacting.

\end{abstract}


 \section*{Introduction}

Suggestions have appeared in the literature that cold dark matter
(CDM) may be in the form of a Bose-Einstein condensate (BEC),
including axionic and other forms of CDM. This has important
implications for the physics of structure formation, notably at
small scales where one expects significant deviations from the more
standard CDM, due to the superfluidity exhibited by BECs. In fact, a
prime motivation for considering self-interacting BECs for CDM was
their ability to produce galactic halos with constant density cores,
see e.g. \cite{goodman} and \cite{peebles}. The corresponding
profiles may then better agree with observed rotation curves of
dwarf and LSB galaxies, see e.g. \cite{BH}, \cite{BMS}. The inner
structure of halos may also depend on the angular momentum
distribution of the CDM particles. Laboratory BECs are known to
develop vortices when rotated with a sufficient angular velocity.
\cite{DS} report that certain fine-structure in the observed inner
mass distribution of the Milky Way can be explained only if the
infalling dark matter particles from which such galactic halos
formed had a net overall rotation, causing a 'tricusp' caustic ring
of dark matter in that case. For standard, non-interacting CDM
models, however, one expects infall to be \textit{irrotational},
while \cite{SY} argue that axionic dark matter, as a BEC, may form
vortices leading to net overall rotation. As such, they suggest,
Milky-Way observations may already have detected the signature of
axionic CDM. Vortices have also been postulated for strongly-coupled
BECs involving ultralight scalar particles (\cite{SM}). The question
of whether an angular velocity sufficient to create vortices occurs
in BEC dark matter cosmologies has not yet been answered, however.
We address this point here by calculating the critical angular
velocity for vortex creation in a simple model of BEC/CDM galactic
halos and comparing the result with the angular velocity expected
from cosmological N-body simulations of CDM.

\section*{The Model}

We shall use an energy argument to derive the critical angular
velocity for vortex creation in a rotating, self-gravitating BEC
halo by finding the angular velocity above which the energy is
lowered by the presence of a vortex. For the unperturbed equilibrium
state, we model the BEC halo as an oblate Maclaurin spheroid - a
homogeneous ellipsoid of mass density $\rho=m n$, with semi-axes
$(a,b,c)$ along $(x,y,z)$ such that $a=b>c$, of total volume $V =
4\pi a^2c/3$, uniformly rotating with angular velocity
$\mathbf{\Omega}=\Omega \mathbf{\hat{z}}$, with gravitational
potential
 $\Phi(r,z) = \pi G \rho \left(A_1(e)r^2 + A_3(e)z^2\right)$
 in cylindrical coordinates
 $(r,z)$. The functions $A_1(e), A_3(e)$ depend on the
 excentricity $e$, defined via $e = \sqrt{1-(c/a)^2}$, and there is
   a family of solutions parameterized by $\Omega/\Omega_G$, or, equivalently, $e$, related
   to each other according to (see \cite{BT})
 \begin{equation} \label{macomega}
  \left(\frac{\Omega}{\Omega_{G}}\right)^2 = 2\left[A_1(e) -
  (1-e^2)A_3(e)\right],
  \end{equation}
   where
     $\Omega_{G} \equiv \sqrt{\pi G \rho}$
   is a characteristic gravitational angular frequency.
    For $\Omega = 0$, $e=0$, while $e$ must not exceed $0.9529$
   (\textit{ibid.}).

We describe these self-gravitating BEC halos of ellipsoidal shape
with varying degrees of rotational support by self-consistently
coupling the Gross-Pitaevskii (GP) equation of motion for the
complex scalar wavefunction $\psi(\mathbf{r},t)$ to the Poisson
equation, where $|\psi|^2(\mathbf{r},t) = n$, the number density of
particles of mass $m$:
 \begin{equation} \label{gp}
 i\hbar \frac{\partial \psi}{\partial t} = -\frac{\hbar^2}{2m}\Delta \psi + (m\Phi + g|\psi|^2 -
 \mu)\psi,~~
  \Delta \Phi = 4\pi G m |\psi|^2.
   \end{equation}
 We assume that halos are comprised of $N$ particles in the condensed state
  described by $\psi$, and so
    $\int_{\mathbb{R}^3} |\psi|^2 = N$,
 which determines the GP chemical potential $\mu$. BEC/CDM, like standard CDM, is assumed to interact so weakly with
other matter and radiation, once its abundance is fixed in the early
universe, that we can neglect all other, non-gravitational
couplings. However, BEC dark matter is \textit{self-interacting}, as
described by an effective interaction potential $g|\psi|^4/2$
  with coupling constant
    $g = 4\pi \hbar^2 a_s/m$, where $a_s$ is
   the 2-body scattering length. [Since BECs with
   negative scattering length are not stable in the context described here, we consider only $g >
   0$.]

 In a frame rotating with velocity $\mathbf{\Omega}$, our system is
 stationary, so the derivative with respect to time in equ.(\ref{gp}) vanishes.
  The equation of motion in this frame is then given by equ.(\ref{gp}) with
an additional operator $V_{rot} = -\mathbf{\Omega}\cdot
 \mathbf{L}$ on the right-hand-side, where $\mathbf{L} = -i\hbar
 \mathbf{r}
 \times \nabla$, and it is also understood that the respective
 variables and quantities are in the new frame.
Incompressible Maclaurin spheroids are approximate solutions of this
system of equations.\footnote{Uniform density is a more realistic
approximation for BEC halos than for the cuspy halos of standard
CDM; equ. (\ref{gp}) - with or without rotation - favours a flat
core over the cuspy $r^{-1}$-profile found by N-body simulations of
standard CDM.} Such stationary systems can then be studied via the
corresponding GP energy functional, given by
  \begin{equation} \label{energie}
   \mathcal{E}[\psi] = \int_{\mathbb{R}^3} \left[\frac{\hbar^2}{2m}
 |\nabla \psi|^2 + \frac{m}{2}\Phi |\psi|^2 + \frac{g}{2}|\psi|^4 + i\hbar \psi^* \mathbf{\Omega}
  \cdot (\mathbf{r}
 \times \nabla \psi)\right].
 \end{equation}

We shall use this equation to determine at which angular velocities
the presence of a vortex is energetically favoured.
   To this aim,
 we decompose the wave function $\psi$ into a vortex-free part and a part which carries the
vorticity, thus splitting the energy functional in
equ.(\ref{energie}) so as
  to compare the respective energy contributions, with or without vortices, more
easily. To derive an analytical result with as much generality as
possible, we consider the following ansatz. The vorticity-part is a
$d$-quantized straight vortex, modeled as a funnel-shaped tube along
the rotation-axis, with core radius $s$, where the density outside
of the core
 is given by the unperturbed profile of the Maclaurin spheroid, whereas the density in the vortex core region
 drops to zero at the center.

\section*{Results}

  When our wave function ansatz is inserted in
  equ.(\ref{energie}), there is a critical angular velocity $\Omega_c$ above which the energy is lowered
  by the presence of a vortex. Since our ansatz leads to an energy greater than or equal to
  that which would result if the 'real' wave function were used in place of the ansatz, this means
  vortex creation is energetically favoured, in general, if $\Omega > \Omega_c$. The general expression
 for $\Omega_c$ (for brevity not shown here; for derivation see \cite{RS}) depends
 on the core radius $s$, which is of the order of the healing
 length $\xi$ defined by equating the energy
 contributions from the kinetic term and
 self-interaction, so
   $\xi^2 = \hbar^2/(2\rho g)$.
In the strongly-coupled regime, as $g \to \infty$, $\xi \to 0$ and
$s \to 0$, respectively, and the critical angular velocity has a
leading-order term which diverges logarithmically,
  $\Omega_c \simeq \Omega_{QM} d \ln \frac{a}{s}$,
 just as for laboratory condensates (see e.g. \cite{lundh}). Here we define $\Omega_{QM} \equiv
 \hbar/(ma^2)$,
   the characteristic angular frequency such that every particle contributes an
   amount $\hbar$ to the total angular momentum of the uniformly rotating Maclaurin spheroid ($|\mathbf{L}| = N\hbar$).
   On the other hand, if $s$ or $\xi$
   are comparable to the system size $a$, this logarithm
 is subleading.
  We note that $\Omega_c$ is a monotonically increasing
  function of the winding number $d$, so we restrict our consideration to the lowest
  $\Omega_c$, for which $d=1$.
 Then, the
 core radius can be replaced by the healing length (see \cite{PS}), and
 the critical angular velocity (in units of $\Omega_{G})$ becomes
  \begin{equation} \label{omega1}
\frac{\Omega_c}{\Omega_{G}} =
 \frac{\Omega_{QM}}{\Omega_{G}}\left[\ln
 \frac{a}{\xi}+\frac{25}{12}+\frac{\pi}{2}\left(\frac{\Omega_{G}}{\Omega_{QM}}\right)^2
 \left(\frac{A_1(e)}{6}\left(\frac{a}{\xi}\right)^{-4}
 + A_3(e)\frac{1-e^2}{9}\left(\frac{a}{\xi}\right)^{-2}\right)\right].
    \end{equation}

We determine whether a given set of BEC parameters $(m,g)$ makes the
spheroid rotation velocity equal the critical value above which a
vortex forms, $\Omega = \Omega_c$, by combining
equs.(\ref{macomega}) and (\ref{omega1}), for a given $e$. We
observe that
  $\Omega_{QM}/\Omega_{G} = m_H/m$ and $a/\xi = (g/g_{min})^{1/2}$,
   where
  $m_H \equiv \hbar/[a^2(\pi G \rho)^{1/2}]$ and $g_{min} \equiv \hbar^2/(2\rho a^2)$,
  the latter of which depend only on halo parameters.
The solution for a given $e$ is a curve in the
$(m/m_H,g/g_{min})$-plane, for which there is a minimum allowed
value of $g/g_{min} \geq 1$. For each curve, no vortex is allowed
for parameters in the space above the curve. According to
equ.(\ref{omega1}), $\Omega_c$ goes to infinity as $g \to 0$. This
is the case for axion dark matter, for which the coupling
  is so weak as to be effectively zero.

We determine the $e$-values of interest for BEC/CDM halos as
follows. The superfluidity effects of BEC dark matter which
distinguish it dynamically from standard CDM are mostly limited to
the internal structure of our halos, while larger-scale structure
formation is otherwise the same. The latter is responsible for the
tidal torques that give a halo its angular momentum. Cosmological
N-body simulations of the CDM universe show that halos form with a
net angular momentum such that the dimensionless ratio
  $\lambda = L |E|^{1/2}/GM^{5/2}$,
which expresses their degree of rotational support, has values in
the range $[0.01,0.1]$ with median value $0.05$ (see e.g.
\cite{BE}), where $L$ is the angular momentum, $E$ is the halo
binding energy, and $M$ is the total mass. For our spheroids,
$\lambda^2$ corresponds roughly to the ratio
 of rotational kinetic energy to gravitational potential energy, and
 can be expressed in terms of $e$ only,
   \begin{displaymath}
   \lambda =
   \frac{6}{5\sqrt{5}}\frac{\arcsin(e)}{e}\left(\frac{3}{2e^2}-1-\frac{3\sqrt{1-e^2}}{2e\arcsin(e)}\right)^{1/2}.
    \end{displaymath}
In what follows, we take three representative values for $\lambda$,
$(0.01, 0.05, 0.1)$, which correspond to $e = (0.051, 0.249,
0.464)$, respectively. For given values of the halo mass density and
eccentricity, the angular velocity is fixed. For example, a
Milky-Way-sized halo with $M = 10^{12}M_{\odot}$ and a radius of
$R=100$ kpc, where the corresponding Maclaurin spheroid with the
same mean mass density has a semi-axis given by $R =
a(1-e^2)^{1/6}$, has a density of about $10^{-26}$ g/cm$^3$ and
$\Omega \sim 10^{-17}$ rad/s for $\lambda = 0.05$, so
$\Omega/\Omega_{G} \sim 0.18$. For such a halo $m_H \sim 10^{-58}$ g
and $g_{min} \sim 10^{-76}$ erg cm$^3$. In Fig.1, we plot $m/m_H$
versus $g/g_{min}$ for the above $\lambda$-values, independent of
halo size.

For BEC halos, the condition of gravitational equilibrium restricts
the values of $m$ and $g$ to another curve in the
$(m/m_H,g/g_{min})$-plane. In the strongly-coupled regime,
non-rotating BEC halos are just ($n=1$)-polytropes, for which the
size $R$ is related to the BEC parameters according to $R = \pi
[\hbar^2 a_s /(G m^3)]^{1/2}$ (see \cite{BH}),
 which we translate into our
 language as
  \begin{equation} \label{comp}
   1 =
   \left(\frac{\Omega_{QM}}{\Omega_G}\right)^2\frac{\pi^2}{8(1-e^2)^{1/3}}\left(\frac{a}{\xi}\right)^2.
    \end{equation}
     The resulting relationship between $m/m_H$ and $g/g_{min}$ is plotted in Fig.1, as well.
      For CDM halo $\lambda$-values, the
     degree of rotational support is small enough that
     equ.(\ref{comp}) should still be a good approximation.
 There is almost no sensitivity to $e$, so
the respective curves for different $e$ lie on top of each other.
Since the BEC parameters which satisfy (\ref{comp}) are all below
our critical curves for which $\Omega=\Omega_c$, BEC/CDM halos, in
general, \textit{will} typically form vortices.

\begin{figure*}
\centering\includegraphics[angle=270,width=10cm]{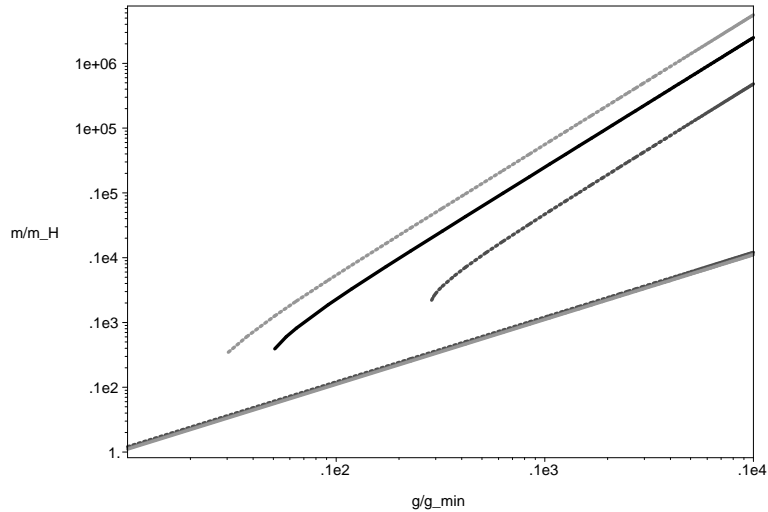}
 \caption{Dimensionless BEC particle mass $m/m_H$ vs. coupling strength $g/g_{min}$:
 critical lines $\Omega = \Omega_c$ for $\lambda = 0.01 ~(e=0.051)$
 (grey-dotted),\newline
 $\lambda = 0.05 ~(e=0.249)$ (black-solid), $\lambda = 0.1 ~(e=0.464)$
 (light grey-dotted); \newline
 BEC halo ($n=1$)-polytropes: lower-most curves (grey-solid) for the same $e$-values} \label{fig}
\end{figure*}

\section*{Conclusions}

While axions, with an effectively zero coupling, apparently do not
form vortices, vortices \textit{will} be created, in general, for
BEC/CDM halos in the strongly-coupled regime. As such, previous BEC
models of halo mass profiles, which do not account for their
presence, should be revised, especially at the centers where vortex
creation causes the density to drop.
\\
\\
\acknowledgements We thank P.Sikivie, and S.Weinberg, E.Komatsu and
other members of the Texas Cosmology Center for stimulating
discussion. This work was supported by the DFG under research unit
FG 960; NSF grant AST 0708176, NASA grant NNX07AH09G and Chandra
grant SAO TM8-9009X.

\end{document}